\documentstyle{article}
\begin{document}

\begin{flushright}
SPbU-IP-96-33
\end{flushright}

\begin{center}
{\Large \bf Chiral Symmetry in algebraic and analytic approaches}\\

\vspace{4mm}
\underline{V. Vereshagin}\\
St.Petersburg State University, St.Petersburg, Russia\\
and\\
Institute for Theor. Physics III, Univ. of Erlangen-Nuernberg,
Erlangen, Germany\\
M. Dillig\\
Institute for Theor. Physics III, Univ. of Erlangen-Nuernberg,
Erlangen, Germany\\
A. Vereshagin\\
St.Petersburg State University, St.Petersburg, Russia\\
\end{center}

\begin{abstract}
We compare among themselves two different methods for the derivation
of results following from the requirement of polynomial boundedness
of tree-level chiral amplitudes. It is shown that the results of the
algebraic approach are valid also in the framework of the analytical
one. This means that the system of Sum Rules and Bootstrap equations
previously obtained with the help of the latter approach can be
analyzed in terms of reducible representations of the unbroken
Chiral group with the known structure of the mass matrix.
\end{abstract}

Novadays it is commonly realised that the Spontaneously Broken
Chiral Symmetry (SBChS) plays an essential role in the hadronization
mechanism provided by the large-distance forces in OCD. The smallness
of the pion mass (compared to the typical hadronic scale of 1 Gev)
made it possible to develop the Chiral Perturbation Theory (ChPT)
\cite{1,2}
(for recent reviews see also
\cite{3,4,5})
allowing one to calculate in a systematic manner the corrections to
the zero-order ("bare") amplitude given by a sum of corresponding tree
graphs. Thus, if the bare amplitudes are known, we can analyze the low
energy processes with the arbitrary high degree of accuracy.

  Unfortunately, the bare (tree-level) amplitudes themselves contain
a lot of free parameters since they depend of infinite number of
basically unknown coupling constants describing the  off-shell
hadron interactions and point-like vertices. Just because of this
reason the current situation looks unsatisfactory: we have an
excellent tool -- ChPT -- for the computation of corrections to
the quantities which we know not so much about.

  A remarkable step in the understanding of the bare couplings was
done by S.Weinberg
\cite{6},
first suggested to use the asymptotic boundedness requirements for
the chiral amplitudes  describing the forward scattering of
massless pions on arbitrary targets (the list of the corresponding
refs. can be found, e.g., in
\cite{7} ;
the paper
\cite{8}
should be added to this list). It was shown that the SBChS -- despite
of its dynamical origin -- manifests itself in a customary algebraic
manner. This means that it is possible to classify the states with a
given helicity according to linear reducible representations of
unbroken Chiral group, the mass matrix being constructed as a sum of
chiral scalar and 4-th component of chiral vector.

  A variety of interesting results derived by many authors in the
framework of Weinberg's scheme clearly demonstrates both the
importance and the fruitfulness of the asymptotic requirements.

  There are known two technically different approaches to the
derivation of results from those requirements.

  Weinberg's original approach
\cite{6}
(later on called algebraic) consisting of the group theoretical
analysis of mass (squared) and coupling matrices, shows both attractive
features and serious limits. One of the most attractive features of
this method is that it allows one to work from the very beginning with
purely algebraic structures. However, it is applicable only for the
case of
\underline{forward}
 scattering of
\underline{massless}
pions; its feasibility is limited on a small number of partial waves
taken into account. Moreover (as stressed by Weinberg himself
\cite{6}),
from the mathematical point of view this method is only valid if the
number of resonances is finite.

  A different approach (first suggested in
\cite{9})
has been recently formulated in its final form in
\cite{7}.
Later on we call this approach analytic. It takes advantage of the
analytic structure of a tree-level amplitude, this structure
being defined by the requirements of meromorphy and polynomial
boundedness. In this method there is no necessity to require that pion
is massless. Also, it is not restricted to forward processes only: the
nonforward ones can be analyzed on the same ground. The method does not
contain any limitations on the number of resonances or partial waves
included. Moreover, it follows from the results of ref.
\cite{7}
that, to avoid self-contradictions in the case if a partial wave with
$l \geq 2$
(or, the same, the resonance with a spin
$J \geq 2$)
is taken into account, one has to include also the full tower of higher
spin resonances  with couplings and masses strongly restricted by a
certain infinite set of self-consistency (bootstrap) equations. So,
compared to the algebraic approach, the analytic one looks more general,
since it contains the algebraic method as a particular case. Moreover, it
contains also the built-in mechanism guaranteing the absence of mutually
contradicting results in those cases when the algebraic method is applied
for the simultaneous analysis of direct- and cross-channel processes.

  However, as a serious drawback, the problem of algebraization of the
system of bootstrap equations -- which is clearly absent in the algebraic
approach -- looks rather complicated in the analytical framework. Thus,
it is interesting to ask the question if it is possible to get
Weinberg's results from the analysis of the corresponding part of a
system of bootstrap equations derived with a help of the analytic method.

  Below we give the positive answer to this question along with the general
outline of the corresponding proof. A more detailed consideration will be
published elsewhere.

  Let us first recall some key points of the analytic approach
(for more details see
\cite{7}).
This approach is based on the following postulates:\\
1. The tree-level amplitude of a given binary process is a meromorphic
function in a space of 3 dependent complex Mandelstam variables
$s, t, u$
($s+t+u=\sum_{i=1}^4 {m_i}^2)$.\\
2. In accordance with the crossing symmetry requirement, this amplitude
describes all three channels of a process under consideration.\\
3. The principal part of the amplitude is completely determined by the
formally written tree-level expression following from the "naive"
(unitary) Feynman rules (no ghosts, no tachions!).\\
4. As a function of every
\underline{one}
complex variable
$x$
(considered as the CMS energy squared in a given channel) this amplitude is
polynomially bounded at zero value of the corresponding momentum transfer,
the degree of a polynomial being dictated by the value of relevant Regge
intercept (or, equivalently, by experiment).

  The main technical tool used in the analytic approach is provided by the
Cauchy method. This method allows one to construct the
\underline{convergent}
power fraction expansion for the meromorphic function
$f(z)$
of one complex variable
$z$
with a given principal part, this expansion being coordinated with the
asymptotic condition of the form
\begin{eqnarray}
\max_{z \in C_n}\left|\frac{f(z)}{z^{N+1}}\right|
\stackrel{n \to \infty}{\longrightarrow} 0  \;,
\label{1}
\end{eqnarray}
where
$C_n$
denotes a smooth contour (for definiteness, a circle) which contains inside
of it the first
$n$
poles (
$|p_{i+1}|> |p_i| , p_i$
is the
$i$-th
pole position,
$r_i$ --
the corresponding residue;
$p_i$
being enumerated in the order of increasing modulo)
and does not contain any other poles. The minimal integer number
$N$
in
(\ref{1})
shows the degree of the bounding polynomial.

  With the condition
(\ref{1})
the Cauchy method gives:
\begin{eqnarray}
f(z)= \sum_{p=0}^N f^{p}(0)\frac{z^p}{p!}+\sum_{i=1}^{\infty}
\left( \frac{r_i}{z-p_i} - {\Pi}_i (z) \right)  .
\label{2}
\end{eqnarray}
Here
${\Pi}_i (z)$
stands for the so-called correcting polynomial: it is nothing but the sum
of the first
$N$
terms of the power series expansion of
$r_i / (z-p_i)$
around the point
$z=0$.
The necessity in correcting polynomials is caused by the convergency
condition. In fact,
the eq.
(\ref{2})
provides a special case of the general theorem by Mittag-Leffler.

  In a framework of the analytic method one works with the amplitude
written in the form
(\ref{2}).
In the case of forward process all the parameters appearing in
(\ref{2})
(namely,
$ f^{(0)}(0), r_i , p_i $ )
are just constants; otherwise, they should be considered as the functions
of the momentum transfer. The form
(\ref{2})
is convenient because it does not contain any unwanted terms breaking the
allowed asymptotic regime, thus from the very beginning there is no need
to control the correctness of the asymptotics.

  This is in strong contrast to the algebraic method, where results are
derived from the condition of cancellation of just those "unwanted"
terms which increase too rapidly with energy but appear
\underline{necessarily}
in the scattering amplitude if the Lagrangian is written in terms of
covariant derivatives (see
\cite{6}).
In this approach one calculates the
\underline{lowest}
Laurent series expansion coefficient which breaks the allowed asymptotic
regime and then requires of it to be zero.

  At the first glance two approaches described above look quite different.
Indeed, since the "naive" Laurent series expansion is only applied for
the study of asymptotics if the number of poles is finite, it looks
impossible to use it for the analysis of cross-conjugated processes where
the infinite set of resonances is required to provide the correct
asymptotic behavior in both channels simultaneously (see
\cite{7}).
This, in turn, could mean that the results of early papers
\cite{10},
based on the calculation of the Laurent coefficients, have no ground. At
the same time, it is possible to show that the Sum Rules given in
\cite{10}
can be also derived with a help of the form
(\ref{2}).
So, it looks that the conclusions obtained in a framework of the
algebraic method remain also valid with respect to the results of the
analytic approach.

  To check (and prove) this idea
\cite{11}
let us consider the case
$N=0$
in
(\ref{1}) and (\ref{2})
(just in order to simplify the corresponding formulas). In this case from
the integral form of the asymptotic condition
(\ref{1})
\begin{eqnarray}
\oint\limits_{C_n}^{}\left| \frac{f(z)}{z^2}\,dz \right|
\stackrel{n \to \infty}{\longrightarrow} 0   ,
\label{3}
\end{eqnarray}
it follows that
\begin{eqnarray}
\sum_{i=n+1}^{\infty} \frac{r_i}{p_i^2}
\stackrel{n \to \infty}{\longrightarrow} 0   .
\label{4}
\end{eqnarray}

  Next, let us consider the true Laurent series expansion
\begin{eqnarray}
f(z) = \sum_{- \infty}^{\infty} C_p^{(n)}z^p,\;\; \;\;\; (z\in R_n)   ,
\label{5}
\end{eqnarray}
where
$R_n$
stands for the ring
$|z| \in (|p_n|, |p_{n+1}|)$
and calculate the lowest "unwanted" coefficient
$C_{+1}^{(n)}$.
It is easy to get
\begin{eqnarray}
C_{+1}^{(n)} = - \sum_{i=n+1}^{\infty}\frac{r_i}{p_i^2}   .
\label{6}
\end{eqnarray}
To study the behavior of the amplitude
$f(z)$
at large values of
$|z|$
one can use the expansion
(\ref{5})
and take a limit
$ n \to \infty $.
In this case -- as it follows immediately from the comparison of
(\ref{6})
with
(\ref{4})
-- one obtains the result
\begin{eqnarray}
C_{+1}^{(n)} \stackrel{n \to \infty}{\longrightarrow} 0   ,
\label{7}
\end{eqnarray}
showing that the lowest "unwanted" Laurent coefficient vanishes (along
with all higher ones
$C_i ^{(n)},   i=+2,+3,...$).
This result justifies Weinberg's formal method for the case of infinite
number of resonances. It explains also the complete coincidence of the
systems of Sum Rules derived in
\cite{10}
(from the formal manipulation with Laurent series expansions) with those
following directly from the convergent partial fraction expansion
(\ref{2}).

  Summarizing, we conclude with the statement that the main result
following from the algebraic method is also valid in the framework of the
analytic one: in the chiral limit
$ m_{\pi} = 0 $
the direct channel resonances with a given helicity fall into reducible
representations of unbroken chiral group
$ SU_2 \times SU_2 $,
the mass matrix being constructed as a sum of a chiral scalar and the 4-th
component of chiral vector.

  This work was supported in part by DAAD (visiting Grant for V.V.),
RFBR (Grant 96-02- 18017) and GRACENAS (Grant 95-06.3-13).

\end{document}